# Mesoscopic, Non-equilibrium Fluctuations
# of Inhomogeneous Electronic States in Manganites

V. Podzorov[1], C. H. Chen[2], M. E. Gershenson[1] and S-W. Cheong[1,2]

[1] Serin Physics Laboratory, Rutgers University, Piscataway, NJ 08854

[2] Bell Laboratories, Lucent Technologies, Murray Hill, NJ 07974

By using the dark-field real-space imaging technique of transmission electron microscopy (TEM), we have observed slow 200 Å-scale fluctuations of charge-ordered (CO) phase in mixed-valent manganites under a strong electron beam irradiation. In addition to these unusual fluctuations of the CO phase, we observed the switching-type fluctuations of electrical resistivity in the same sample, which were found to be as large as several percents. Systematic analysis indicates that these two different types of fluctuations with a similar time scale of the order of seconds are interconnected through a meta-stable insulating charge-disordered state. Current dependence of the fluctuations suggests a *non-equilibrium* nature of this slow dynamics.

Recent extensive research on manganites has revealed that coexistence, as well as competition, of the antiferromagnetic charge-ordered insulating (CO) and ferromagnetic metallic (FM) phases result in the extraordinary transport properties of these compounds [1-3]. In the presence of both CO and FM phases, coexisting on a ~ 5000 Å-scale, electronic transport occurs through the percolation network of metallic regions [4,5]. The *1/f* noise, diverging near the Curie temperature, corroborates the percolation nature of the metal-insulator (M-I) transition in manganites [6]. The M-I transition can be drastically influenced by varying internal parameters, such as chemical composition or external parameters, such as magnetic or electric fields and x-ray irradiation [5,7-10]. Colossal magnetoresistance (CMR) in manganites is directly related to the high sensitivity of the M-I transition to these parameters.

The CO pattern in mixed-valent manganites is composed of the $Mn^{3+}$ charge stripes with a charge modulation wavelength of 10-30 Å determined by the $Mn^{3+}$ concentration [11]. Similar charge stripes have been commonly observed when charge carriers are introduced into antiferromagnetic (AF) Mott insulators. Examples include layered cuprates and nickelates [12]. Especially, in the case of cuprates, the dynamic nature of charge stripes and also the role of dynamic charge stripes in superconductivity have been recently extensively discussed [13]. Various scattering experiments on manganites, however do not show any dynamic nature of charge stripes, indicating that the charge stripes in manganites are well frozen at low temperatures [14].

On the other hand, intriguing dynamics of electronically inhomogeneous manganites has been revealed in the so-called telegraph noise in the resistivity $r$ [15]. The thin film manganites demonstrate switching between low-$r$ and high-$r$ states with the change of $r$ of the order of 0.02 %. This switching was interpreted as equilibrium thermal fluctuations between two electronic states.

In our study of the temporal evolution of $r$, we found the giant switching of $r$ up to 7 % with the time scale of ~ 1 s in bulk samples of low-$T_C$ manganites. Furthermore, by using transmission electron microscopy (TEM), we have observed that a charge-ordered (CO) domain would gradually break up into smaller charge-disordered (CD) domains (~ 200 Å in size) with increasing incident electron beam intensity. At a fixed electron beam intensity of TEM, slow dynamic fluctuations between these two states (CO and CD) are observed, and the time scale of these fluctuations is similar to those observed in $r$-fluctuations. Both types of fluctuations, observed in TEM and in $r$, are very sensitive to the measuring current, suggesting a non-equilibrium nature of these processes. We argue that these fluctuations originate from meta-stability of the charge-disordered insulating (CD-I) phase, which is an intermediate state between relatively stable CO and FM states.

Transport and TEM measurements have been carried out on bulk samples of $La_{0.225}Pr_{0.4}Ca_{0.375}MnO_3$ with dimensions ~ 1.5×1.5×2 mm$^3$. The resistivity was measured with the standard four-probe method over the temperature range $T$ = 4.2-300 K. The long-term stability of $T$ in our transport measurement was better than ~ 1 mK at $T$ ~ 10-300 K. A bias current $I = 10^{-8}$–$10^{-6}$ A was driven through the sample by a low-noise current source with the output resistance much greater than the sample resistance.

The temperature dependence of $r$ is shown in Fig. 1. With cooling below the CO transition ($T_{CO}$ ~ 210 K), $r$ increases sharply until the system undergoes the M-I transition at $T_C$ ~ 40 K. Below 40 K, $r$ remains almost constant. Previous studies have demonstrated that the M-I transition is of a percolation type, and that the large value of $r$ at low $T$ is due to coexistence of metallic and insulating phases. A signature of the percolative M-I transition in low-$T_C$ manganites is a strong thermal hysteresis of $r$ in the interval $T$ ~ 15-100 K. Slow monotonic relaxation of the resistivity from a meta-stable high-$r$ (solid line) to a more stable low-$r$ state (dashed line) is typical when $T$ is fixed in this interval [16].

We observed discrete discontinuities superimposed on the monotonic relaxation of $r$ at fixed $T$ in the range of thermal hysteresis $T \sim$ 35-75 K (see the insets in Fig. 1). The stepwise drops of $r$ (inset 1) were found to be more typical for the high-$r$ state (the solid line). Another type of discontinuities, switching of $r$ back and forth, is typical when the sample is in the stable low-$r$ state (the dashed line). Examples of switching in $r(t)$ back and forth are shown in the insets 2 and 3. Amplitude of the switching near the M-I transition is as large as $\sim$ 7 % (inset 3). Switching has been found to be very sensitive to the measuring current (inset 4). Neither steps nor switching of $r$ have been found outside of the thermal hysteresis interval.

In order to gain further insight into the origin of fluctuating $r$, we have investigated the local structural properties of the same sample by the dark field imaging technique of TEM. In this technique, the most intense electron diffraction superlattice spot from the charge ordering is used to form a real-space image of the sample. Since only the superlattice reflections are used, not the main Bragg peaks, the bright contrast in this imaging mode is directly associated with the charge ordered domains. The dark field images reveal coexistence of CO and charge-disordered (CD) domains at low temperatures, consistent with previous investigations. Under low electron beam intensity, it is found that the domains of different types are stable with time at a fixed $T=$ 90 K. However, as the incident beam intensity gradually increases, the CO domains are found to break up into smaller CD domains as shown in Fig. 2. At low beam intensity ($\sim$ 50 mA/cm$^2$ at the sample), the CO domains exhibit relatively uniform contrast with the exception of the curvy anti-phase boundaries (Fig. 2a). Small isolated CD clusters ($\sim$ 200 Å) with dark contrast start to appear at intermediate beam intensities $\sim$ 500 mA/cm$^2$ (Fig. 2b). With further increase of the beam intensity to ~ 20 A/cm$^2$, more CD clusters are nucleated and, eventually, a continuous path of the CD phase is established (Fig. 2c). Note that CD-insulating and CD-metallic phases cannot be distinguished in the dark field TEM studies: both appear as dark regions. The typical CD cluster size is ~ 200 Å with no significant further growth after its nucleation. The whole process is found to be reversible with the change of the electron beam intensity. After reducing the beam intensity to the initial level, we found that most of the CD clusters have reverted back to the CO state except for a few that are strongly pinned (Fig. 2d). Given enough time, even the strongly pinned CD clusters eventually are transformed back to the CO state. Similar behavior of the coexisting CO and CD domains has been observed at other temperatures within the entire hysteresis loop. These observations directly show that the transition from CO to CD states can be induced by a strong electron beam irradiation [17].

One of the most intriguing observations of the electron-beam-induced CO-CD transition is the slow fluctuations of small clusters between the CO and CD states under constant beam intensity of TEM. Figure 3, reproduced from a recorded videotape, shows a sequence of dark field TEM images obtained consecutively from the same area at 2-second time intervals. To show that the small clusters are "popping" in and out of contrast, we have plotted the integrated intensity of a chosen cluster (marked "C" in Fig. 3a) normalized to the integrated intensity of each frame as a function of time (Fig. 4). The cluster is continuously switching back and forth between the CO and CD states with the time scale of ~ 2 s comparable to the time scale of switching in our transport measurements. We observed similar fluctuations at different temperatures well below 90 K (e.g. in the entire hysteresis region), where they were almost temperature independent.

It should be emphasized that the fluctuations between CO and CD phases in the TEM experiment were not detected at low electron beam intensities, and were observed only when the electron beam intensity was high enough. In other words, these fluctuations are a non-equilibrium process, induced by the electron beam of a high enough intensity. Heating by the electron beam can be neglected since the microstructural changes associated with the phase transitions observed by TEM occur at the same temperatures as determined by the transport and magnetic susceptibility measurements. To prevent inelastic scattering and electrons trapping, the sample in TEM experiment was thinned from bulk to its final thickness about 50 nm, and the problem of sample charging was avoided. In the transport measurements, the observed discrete fluctuations of $r$ also exhibit a very strong dependence on the measuring current, which suggests a current-driven non-equilibrium nature of this process. In the measurements of $r$ at fixed $T$, a small variation of the measuring current $\delta I$ = 5-20 % results in changes of the switching rates or could lead to steps in $r(t)$ or even a complete fade-out of switching (Fig. 1, inset 4). With restoring the initial value of $I$, switching usually recovers. Note that $I$ in our transport measurements was well below the level at which heating of the entire bulk sample occurs ($I \sim 1$ µA).

This non-equilibrium, oscillatory behavior of $r$ can be induced by local heating of a small region of the percolation cluster without changing the average temperature of the sample. Significant *local heating* can occur due to highly non-uniform current in an electronically inhomogeneous sample. Indeed, near the percolation threshold, the local current density in a percolation link can be very high. Due to the local increase of $T$, the link undergoes a transition from a low-$T$ conducting to a high-$T$ insulating state, reducing the local current and temperature and, hence, undergoing the transition back to the low-$T$ conducting state. This self-oscillation process causes periodic fluctuations of $r$. The frequency of such oscillations depends on the electrical conductivity of the metallic domains and the thermal conductivity $\kappa$ of the insulating domains. Two-level switching in a spherical particle of radius $r$ can be modeled by the heat balance equation:

$$P\tau - 4\pi r^2 \kappa \Delta T / r = (4/3)\pi r^3 \rho C_V \Delta T.$$

The first term represents the Joule energy released in the particle during time $\tau$. The second term is the energy removed from the particle due to heat conduction. The right-hand side of the equation is the energy necessary to increase the temperature of a particle with the density $\rho$ and specific heat $C_V$ by $\Delta T = T_{local} - T_{average}$. Typical values for $\kappa$, $C_V$ and $\rho$ for manganites at $T \sim$ 50 K correspond to $10^{-2}$ W/cmK, 20 J/moleK and 7 g/cm$^3$, respectively [18]. If a percolation link provides fluctuations of the resistance $\Delta R/R \sim 1$ %, the Joule power $P$ generated in the link can be roughly estimated as $P_0 \Delta R/R$, where $P_0 < 0.1$ mW is the Joule power generated in the entire sample. The exact value of $P$ depends on the structure of the percolation cluster. Assuming that $\Delta T \sim 10$ K is necessary for switching between metallic and insulating states of the domain, the above equation yields the size of the oscillating region $\sim 100$ Å for the experimental $\tau \sim 1$ s.

Earlier experiments have suggested that the CO and FM phases are two states responsible for the M-I transition in electronically inhomogeneous manganites. However, recent scattering experiments on manganites have shown that despite a pronounced M-I transition, the concentration of the CO phase *increases* below the M-I transition, implying that the CO phase is stable at low temperatures, and does not play a major role in the transition [19]. Another insulating phase *without long-range charge order*, rather than the CO phase, should be relevant to the percolation at the M-I transition. It can be called the charge-disordered insulating (CD-I) phase. This CD-I phase arises, most likely, as a result of the accommodation of significant strain induced by the lattice mismatch between the CO and FM phases. Indeed, significant structural deformations at the CO transition induces long-range strain in manganites. Naturally, this CD-I phase is meta-stable at low-$T$, and can be readily converted to one of the stable FM or CO phases by any perturbation, such as an electron beam or measuring current.

Existence of the meta-stable CD-I phase at low temperatures sheds light on the origin of the fluctuations in both transport and TEM experiments. Switching in transport measurements reflects fluctuations between the conducting FM state and one of the insulating states – CD-I or CO. However, metastability of the CD-I phase and robustness of the CO phase at low $T$ suggest that the insulating phase participating in $r$-switching is CD-I, rather than CO. On the other hand switching observed in TEM reveals fluctuations between the CO state and one (or both) of the charge-disordered states – CD-I or FM. The dark field TEM imaging technique does not allow to determine whether the latter state is CD-I or FM. Metastability of the CD-I phase, however, suggests that the fluctuations in TEM occur between the CO and CD-I states, rather than between CO and FM states. Therefore, both types of fluctuations likely involve the charge-disordered state. These two fluctuation phenomena, being observed under very different experimental conditions, represent different processes taking place in the same solid and are interconnected through the metastable CD-I state.

In summary, we found that slow fluctuations observed in the resistivity and dark-field images in TEM occur as a result of non-equilibrium processes in a system involving two stable phases - insulating charge-ordered and conducting ferromagnetic, and a meta-stable charge-disordered insulating phase. The meta-stable CD-I phase, which retains much of the characteristics of the high temperature paramagnetic phase, also plays an important role in the CMR effect.

This work was partially supported by the NSF Grant No. 9802513. We thank M. Uehara for sample preparation, Lixin He for computer analysis and V. Kiryukhin and M. Weissman for helpful discussions.

**Figure captions**

**Fig. 1**   Temperature dependence of $r$ (solid line - cooling, dashed line - heating) for a bulk sample $La_{0.225}Pr_{0.4}Ca_{0.375}MnO_3$. Inset 1 shows $r(t)_{T=60K}$ after the sample was cooled from 300 K to 60 K within an hour. Inset 2 shows $r(t)_{T=60K}$ after a subsequent long relaxation for ~ 10 hours at $T = 60$ K. Inset 3 shows $r(t)_{T=42K}$ after the sample was cooled down to 4.2 K and then heated up to 42 K. Inset 4 shows $r(t)_{T=52K}$ for different measuring currents after the sample was cooled down to 52 K.

**Fig. 2**   TEM dark field images obtained at fixed $T = 90$ K (after cooling from 300 K) with different electron beam current densities. The reversible changes of CO are observed with increasing intensity of the electron beam irradiation. Bright (dark) speckles correspond to charge-ordered (disordered) domains. The estimated electron beam current densities are 50, 500, 20000, and 50 $mA/cm^2$ for (a), (b), (c), and (d) correspondingly. The size of each panel is 430nm x 360nm.

**Fig. 3**   Sequence of dark field TEM images obtained with 2 seconds time intervals, which shows the domain fluctuations between CO and CD states at a fixed temperature (90 K) and a constant intensity of the incident electron beam (~ 500 $mA/cm^2$). A few clusters are marked as A, B, and C to highlight their fluctuations with time. The size of each panel is ~ 100nm x 100nm.

**Fig. 4**   The normalized integrated intensity of the cluster marked "C" in Fig.3 as a function of time. The cluster fluctuates back and forth between the CO and CD states with the time scale of ~ 2 s. The main error in this plot is dominated by the ability to choose exactly the same area of the sample from frame to frame for the integrated intensity.

Fig. 1

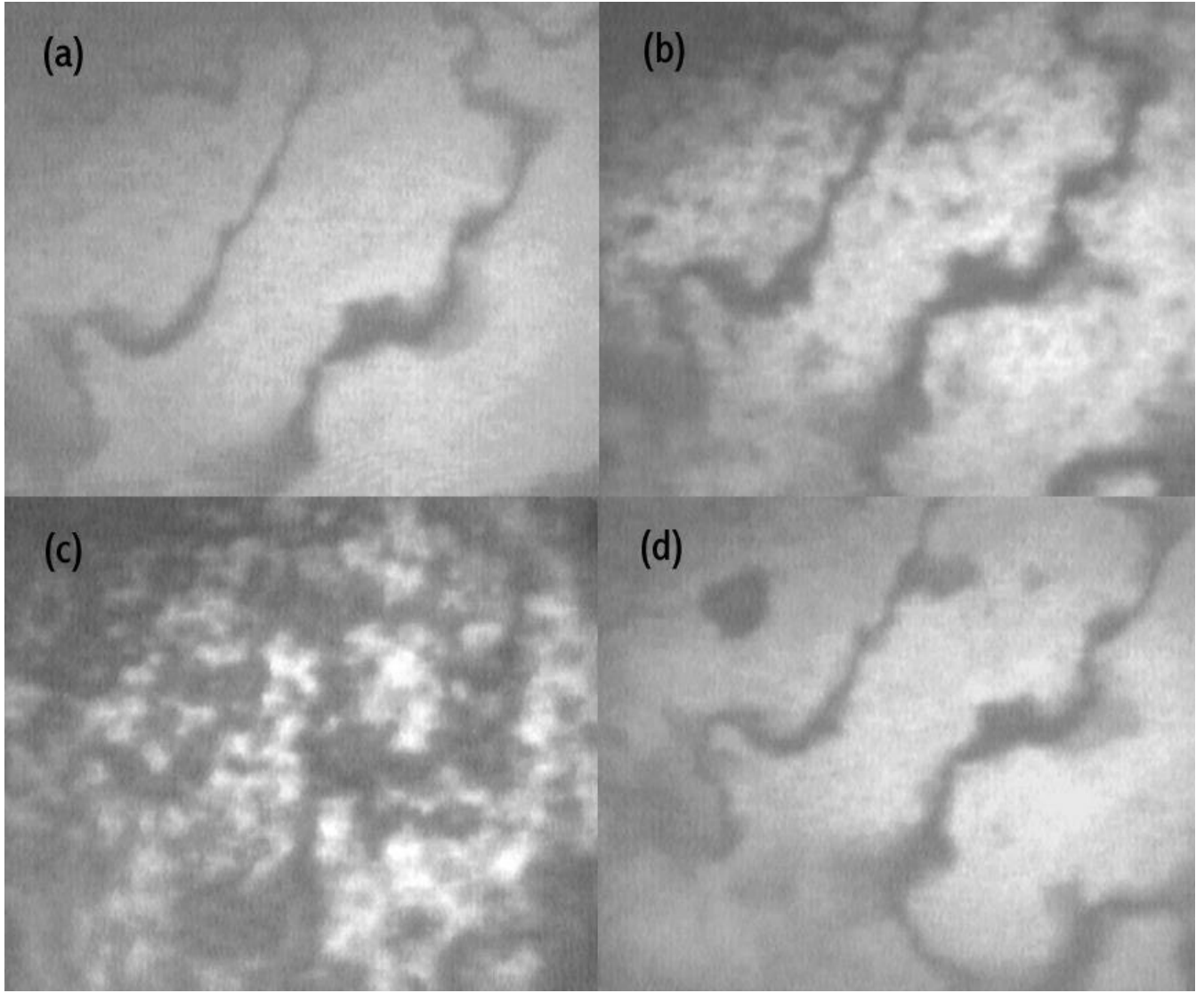

Fig. 2

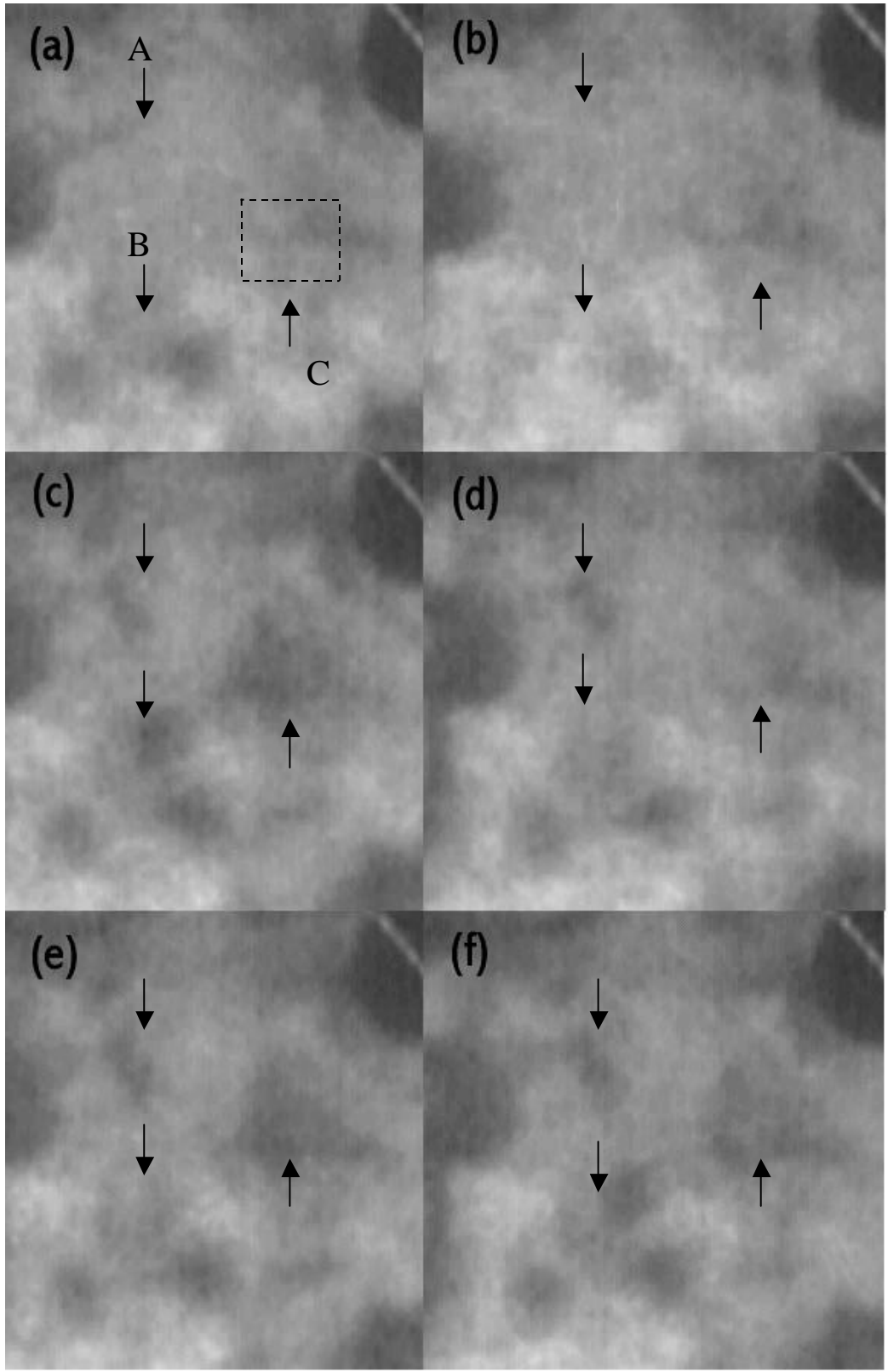

Fig. 3

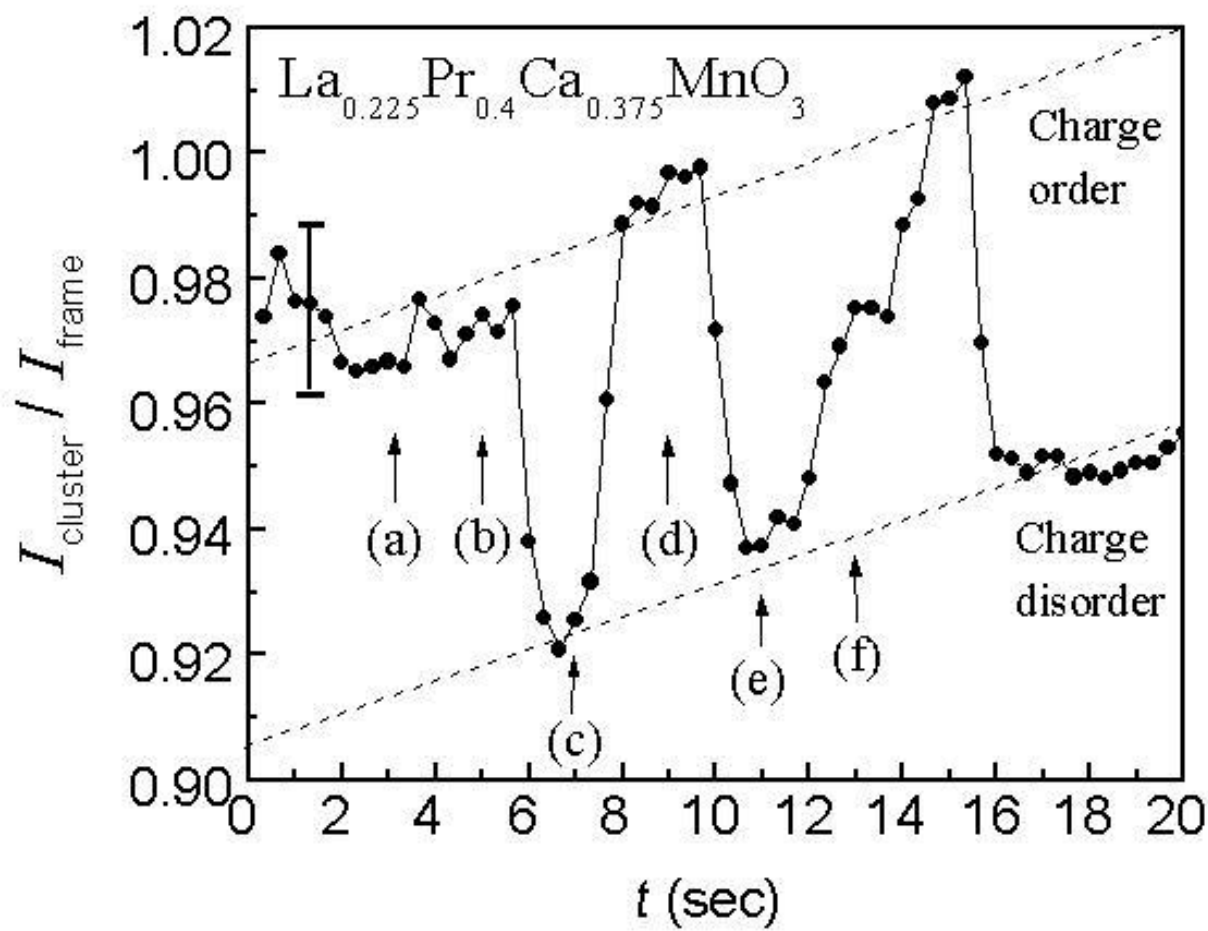

Fig. 4